\documentstyle[12pt,fleqn,epsfig]{article}
\begin{document}

\begin{flushright}
CERN-TH/99-14
\end{flushright}

\setcounter{footnote}{0}
\renewcommand{\thefootnote}{\alph{footnote}}


\begin{center}

{ \Large \bf 
Universal Pion Freeze-out Phase-Space Density\\[0.5cm]
}

D. Ferenc$^{a,b}$, U. Heinz$^{a,c}$, B. Tom\'a\v sik$^a$, 
U.A. Wiedemann$^{a,d}$\\[0.3cm] 
  
{\it $^a$Institut f\"ur Theoretische Physik, Universit\"at Regensburg,\\
D-93040 Regensburg, Germany\\
$^b$Physics Department, University of California, Davis, CA 95616-8677,USA\\
$^c$Theory Division, CERN, CH-1211 Geneva 23, Switzerland\\
$^d$Physics Department, Columbia University, New York, NY 10027, USA}\\[0.4cm]

and J.G. Cramer\\[0.3cm]

{\it Nuclear Physics Laboratory, University of Washington, 
Seattle, WA 98195, USA} 

\end{center}

\vskip 0.2cm

\noindent{\bf Abstract:}
{
Results on the pion phase-space density at freeze-out in
sulphur-nucleus, Pb-Pb and $\pi$-p collisions at the CERN SPS 
are presented. All heavy-ion reactions are consistent with 
the thermal Bose-Einstein distribution $f=[\exp(E/T)-1]^{-1}$ at $T\sim$
120 MeV, modified for radial expansion. $\pi$-p data are also
consistent with $f$, but at $T\sim$ 180 MeV and without radial flow.
}

\vspace{0.2cm}

\begin{center}
\today
\end{center}

\setcounter{footnote}{0}
\renewcommand{\thefootnote}{\arabic{footnote}}

{\it Introduction.}
In ultrarelativistic heavy-ion collisions, the pion freeze-out
phase-space density determines the importance of multiparticle
pion correlations and of dileption production from $\pi^+\pi^-$
annihilation. G. Bertsch \cite{Bertsch} suggested a way of estimating
this quantity by combining measurements of one-particle momentum
spectra with two-particle correlations, thereby testing the
thermal equilibrium of the pion source created in the collision.
In local thermal equilibrium at temperature $T(x)$ the pion energy 
distribution is given by the Bose-Einstein function
 \begin{equation}
 \label{1}
   f(x,p) = {1\over e^{p\cdot u(x)/T(x)}-1},
 \end{equation}
where $u(x)$ is the 4-velocity of the local rest frame at point $x$ in 
the observer frame. The coordinate space average of this function 
is the quantity to be measured:
 \begin{equation}
 \label{12}
   \langle f \rangle(p) = 
   \frac{\int f^2(x,p)\, p^\mu d^3\sigma_\mu(x)} 
        {\int f(x,p)\, p^\mu d^3\sigma_\mu(x)} .
 \end{equation}
Here $d^3\sigma(x)$ is the normal vector on a space-like space-time
hypersurface $\sigma(x)$. According to Liouville's theorem, $\sigma$
is arbitrary as long as its time arguments are later than the time
$t_{\rm f}({\bf x})$ at which the last pion passing the surface at
point ${\bf x}$ was produced. 

If the measured single-particle $p_T$-spectrum is parameterized by
an exponential with inverse slope parameter $T_{\rm eff}(y)$,
 \begin{equation}
 \label{3}
   {dn^- \over dy\, p_T dp_T} = 
   {dn^-\over dy}\, {1\over T^2_{\rm eff}(y)} \,
   \exp \left( -\frac{p_T}{T_{\rm eff}(y)} \right) \, ,
 \end{equation}
and the two-particle correlation function by a Gaussian \cite{cross-term},
 \begin{equation}
 \label{4}
   C(q;p_T,y) = 1 + \lambda(p_T,y) \exp\left(
   -\sum_{i=o,s,l}R_i^2(p_T,y) q_i^2 
   - 2 R_{ol}^2(p_T,y) q_o q_l \right),
 \end{equation}
where the subscripts $o,s,l$ refer to the usual out-side-long
coordinate system and $\lambda(p_T,y)$ accounts for unresolvable
contributions from long-lived resonances \cite{WH97}, one obtains for
the spatially averaged phase-space density at freeze-out \cite{AGS,WH99}  
 \begin{equation}
 \label{5}
   \langle f \rangle (p_T,y) = \sqrt{\lambda(p_T,y)}\, 
   {{dn^-\over dy}\, {1\over 2\pi T^2_{\rm eff}(y)} \, 
    e^{-p_T/T_{\rm eff}(y)} 
    \over 
    \pi^{-3/2}\, E_p\, R_s(p_T,y) 
    \sqrt{R^2_o(p_T,y) R^2_l(p_T,y) - R^4_{ol}(p_T,y)}}\, .
 \end{equation}
Here $E_p{=}\sqrt{m^2{+}{\bf p}^2}{=}m_T\cosh y$, with 
$m_T=\sqrt{m^2+p_T^2}$. The numerator (with experimental input 
$dn^-/dy,T_{\rm eff}(y)$) gives the momentum-space density at freeze-out 
while the denominator (involving the measured two-pion Bose-Einstein 
correlation radii $R_s, R_o, R_l, R_{ol}$) reflects the space-time 
structure of the source at freeze-out and can be interpreted as its 
covariant homogeneity volume for particles of momentum $p$. The factor 
$\sqrt{\lambda}$ ensures \cite{AGS} that only the contributions of 
pions from the decays of short-lived resonances, which happen close to 
the primary production points, are included in the average phase-space 
density. 

We have calculated $\langle f \rangle (p_T,y)$ for the S-S, S-Cu,
S-Ag, S-Au, S-Pb and Pb-Pb data from the experiments NA35~\cite{NA35},
NA49~\cite{Hari,NA49}, and NA44~\cite{NA44}, and for the $\pi$-p data
from the NA22 experiment~\cite{NA22} (all at the CERN SPS). The
projectile energies were 200 GeV per nucleon in S-nucleus collisions,
158 GeV per nucleon in Pb-Pb collisions and 250 GeV in $\pi$-p
collisions, which correspond to projectile rapidities 6, 5.8 and 8.2,
respectively. Our results will also be compared with the average
phase-space density in Au-Au collisions at projectile momentum 10.8
GeV/c, published by the E877 collaboration at the AGS~\cite{AGS}. In all
cases the analysis was done in the LCMS (longitudinally comoving 
system) where the longitudinal momentum of the pion pair vanishes.

\vskip 0.3cm

{\it Calculation of the phase-space density.}
The experimental input into Eq.~(\ref{5}) is partly incomplete. This
concerns particularly the intercept parameter $\lambda$ which was
measured in all the considered experiments but not always published in
the final corrected form. However, the necessary information is still
available, and in the following we explain how we used it. The
following experimental effects are responsible for the uncertainties
in the measurement of the $\lambda$ parameter:

\renewcommand{\labelenumi}%
             {(\roman{enumi})}
\begin{enumerate}

\item
Finite momentum resolution reduces the correlation intercept, since it
leads to a smearing or widening of the correlation peak.

\item
Corrections for Coulomb repulsion of like-sign pions play an important
role in the Bose-Einstein correlation analysis, in particular in the
measurement of $\lambda$. Certain sophisticated Coulomb correction
methods have been applied~\cite{NA35,NA49,NA44,Hiroshima,Coulomb},
but not to all the data sets, as we shall discuss below.

\item
If pions are not positively identified, as is the case in NA35 and
NA49, the direct-pion sample is contaminated with kaons, converted
electrons, protons and other particles. NA35 and NA49 have
performed a contamination correction based on detailed Monte Carlo
simulations~\cite{NA35}, but unfortunately the resulting corrected
results for $\lambda$ have not been directly published. However, since
the contamination levels were published, as well as the uncorrected
results for $\lambda$, we were able to estimate the corrected values
ourselves. For example, in the NA49 Pb-Pb collisions, in the rapidity
interval 3.4$<y<$3.9, the fraction of pure $\pi^- \pi^-$ pairs is 
$x \simeq 55$\%~\cite{Hari}, and the uncorrected measured
$\lambda_{meas}$ is between 0.4 and 0.5~\cite{Hari}. The
contamination-corrected value is then $\lambda_{corr} =
\lambda_{meas}/x \sim$ 0.73-0.91. 

\item
Pions originating from weak decays also contaminate the direct pion
sample when the decay takes place unresolvably close to the main
interaction point. In the rejection of decayed pions, the experiments
NA35 and NA49, with continuous tracking detectors covering the region
close to the target, are in an advantage over the NA44 spectrometer
with tracking devices placed only far away from the target. Since
there are more hyperons ($\Lambda$, $\Sigma$, $\Xi$) than antihyperons
in the fireball which produce more negative than positive pions, the 
measurement of $\lambda$ with positive pions is less affected and should 
be more reliable. Note that NA44 indeed reported~\cite{NA44} 
significantly different results: $\lambda^{--}\approx 0.52 \pm 0.03$ and
$\lambda^{++}\approx 0.66 \pm 0.04$, the difference being
mostly reproduced in a detector simulation with the RQMD event
generator, and thereby traced to the weak decay asymmetry.

\end{enumerate}

NA44 and NA49 both reported an approximate independence of $\lambda$ on
$p_T$, valid to about $\pm$10\%. Regarding the absolute values, for
S-Pb data NA44 has reported $\lambda^{++}({\rm NA44})\simeq
0.52-0.59$ (with negligible non-pion contamination), consistent 
with the lower half of the NA35 interval, $\lambda^{--}({\rm NA35})
\simeq 0.55-0.7$ (contamination corrected). The latter may be
read out directly from the correlation functions presented in Fig.~2
of Ref.~\cite{NA35}. That figure also provides an insight into the
reliability of the $\lambda$ measurement, since it presents the
evolution of the correlation function through different correction
steps.  

\begin{figure}[*]
\epsfig{file=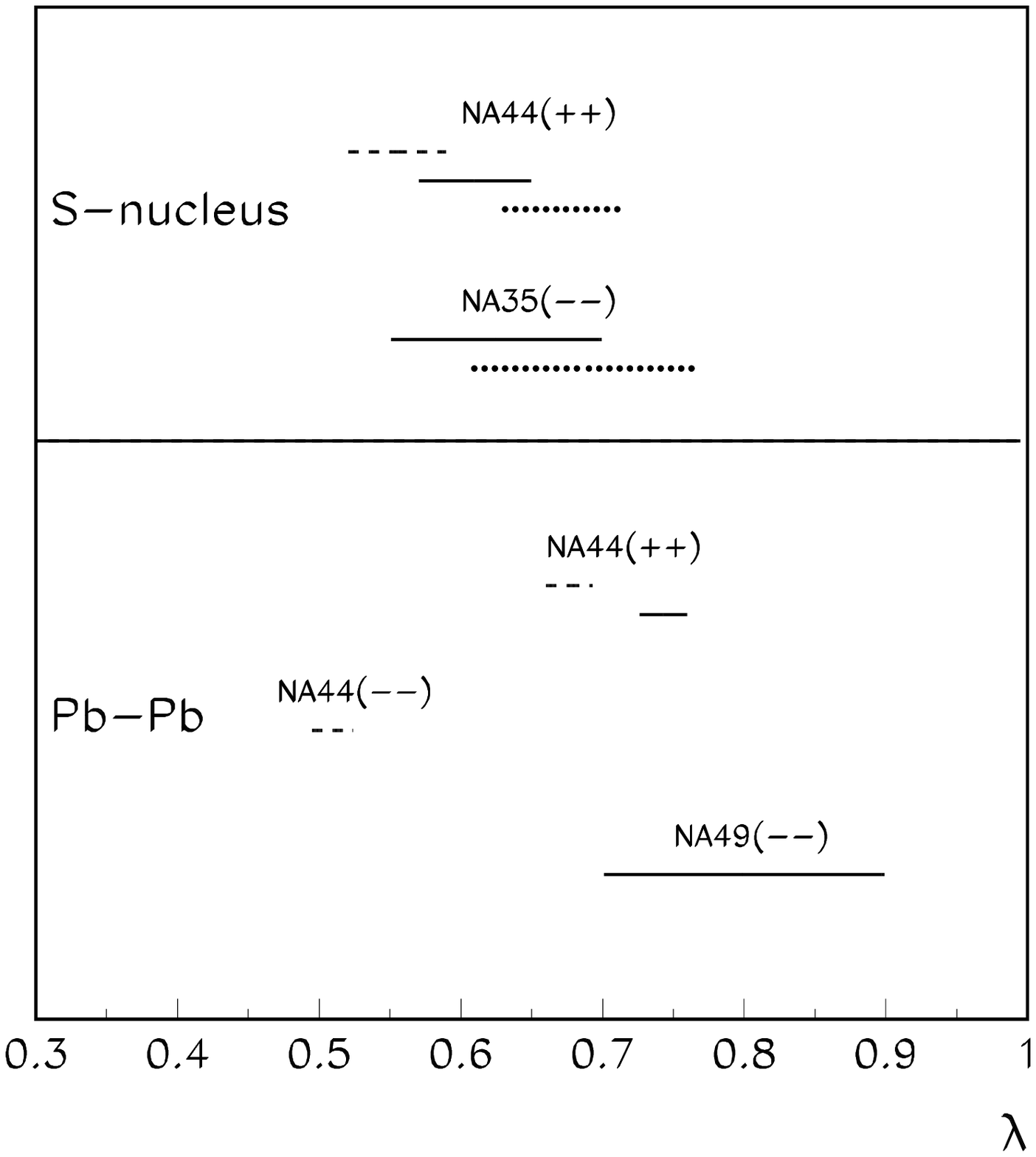,bbllx=0,bblly=0,bburx=500,bbury=500,width=16cm}
\caption{
Correlation strength (intercept parameter) $\lambda$. Dashed lines:
raw uncorrected results. Full lines: contamination corrected results.
Dotted lines: ad hoc 10\% correction for inappropriate Gamov Coulomb
correction. 
}
\label{FPP0}
\end{figure}

In case of Pb-Pb collisions the NA44 results are again consistent with
the lower end of the NA49 interval, see Fig.~\ref{FPP0}:
$\lambda^{--}$(NA49)$\sim$0.7-0.9, obtained with ``our" correction for 
contamination, and $\lambda^{++}$(NA44)$\sim$0.66-0.69, without a
correction for the contamination due to weak decays. Since positive
pions are also contaminated by weak decay products, the NA44 result is
certainly an underestimate, unlike the NA49 result which has been
corrected for all sources of contamination, including those weak
decays which could not be filtered out by vertex cuts. 

If only the contamination by weak decay products were to explain the
difference between $\lambda^{--}$(NA44) and $\lambda^{++}$(NA44), as
suggested by NA44~\cite{NA44}, then the contamination effect itself on
$\lambda^{++}$(NA44) should be quite strong, probably on the order of
10\%. After shifting the NA44 results up by this ad hoc 10\%
correction, as shown in Fig.~\ref{FPP0} by full lines, the NA44 and
NA35/NA49 intervals already fully overlap.

In comparing S-nucleus and Pb-Pb results one has to take into account
also a difference in the Coulomb correction methods. Sophisticated
Coulomb correction methods lead to higher $\lambda$ measurements than
the ``old" Gamov correction. This effect seems to be around 5-15\% for 
Pb-Pb data~\cite{NA49,NA44}, and should be lower for S-nucleus
data~\cite{Coulomb,Hiroshima}, probably 0-10\%, depending on the way
the normalization has been done~\cite{NA35,Hiroshima,Coulomb}. Note
that the Pb-Pb data have been Coulomb-corrected in the appropriate
way, while the S-nucleus data were only Gamov corrected, which might
be the reason for the discrepancy seen in Fig.~\ref{FPP0}. Assuming
that the S-nucleus data indeed need to be corrected by 10\% to account
for the systematical error due to the inappropriate Gamov correction,
one arrives at the results presented by dotted lines in
Fig.~\ref{FPP0} which are in much better agreement with the Pb-Pb
results. Our assumption is, however, rather doubtful and should serve
as an illustration of the systematic uncertainties rather than as a
quantitative result. Note that also the correlation radii should in
principle increase due to a proper Coulomb correction, which would
partly cancel or even reverse the effect of an increased $\lambda$,
see Eq.~(\ref{5}).

It is beyond our ability to check all these points in further detail,
since this would require detailed Monte Carlo simulations of the
detector response, with a realistic event generator as input. It is
quite obvious that with the quality of the presently available data an
accurate determination of the factor $\sqrt{\lambda}$ to be used in
Eq.~(\ref{5}) is not possible. We have therefore proceeded in a rather
pragmatic way, taking the most probable interval for $\lambda$ for all 
the data sets from Fig.~\ref{FPP0}, based on a consideration of the
corrected data presented by full (and to some extent also dotted)
lines in Fig.~\ref{FPP0}. The most likely value for all these data
sets seems to be $\lambda$= 0.7, with a relatively wide error
$\pm$0.1. This value has been used in our analysis.

Fortunately, $\lambda$ enters Eq.~(\ref{5}) under a square root, and
a systematical error of e.g. 15\% for $\lambda$ will result in only
7\% error of the final result, which raises the confidence into our
simplistic approach. 

In the introduction we stated already that we shall use an exponential
function, Eq.~(\ref{3}), to parametrize the $p_T$-spectrum. One
of the reasons for our choice was the general accessibility of
$T_{\rm eff}$ for all the data sets; $T_{\rm eff}$ equals 
$\langle p_T \rangle/2$, a quantity which is usually quoted
together with the data. The best method would be to use directly the
measured $dn^-/d^2p_T$, but most of the data have not been
published in a useful form. The pion rapidity density $dn^-/dy$
was derived from the NA35 and NA49 measurements of the negative hadron
rapidity density $dn^{h-}/dy$, by a simple scaling with a factor
0.9~\cite{Marek}.  

\begin{figure}[*]
\epsfig{file=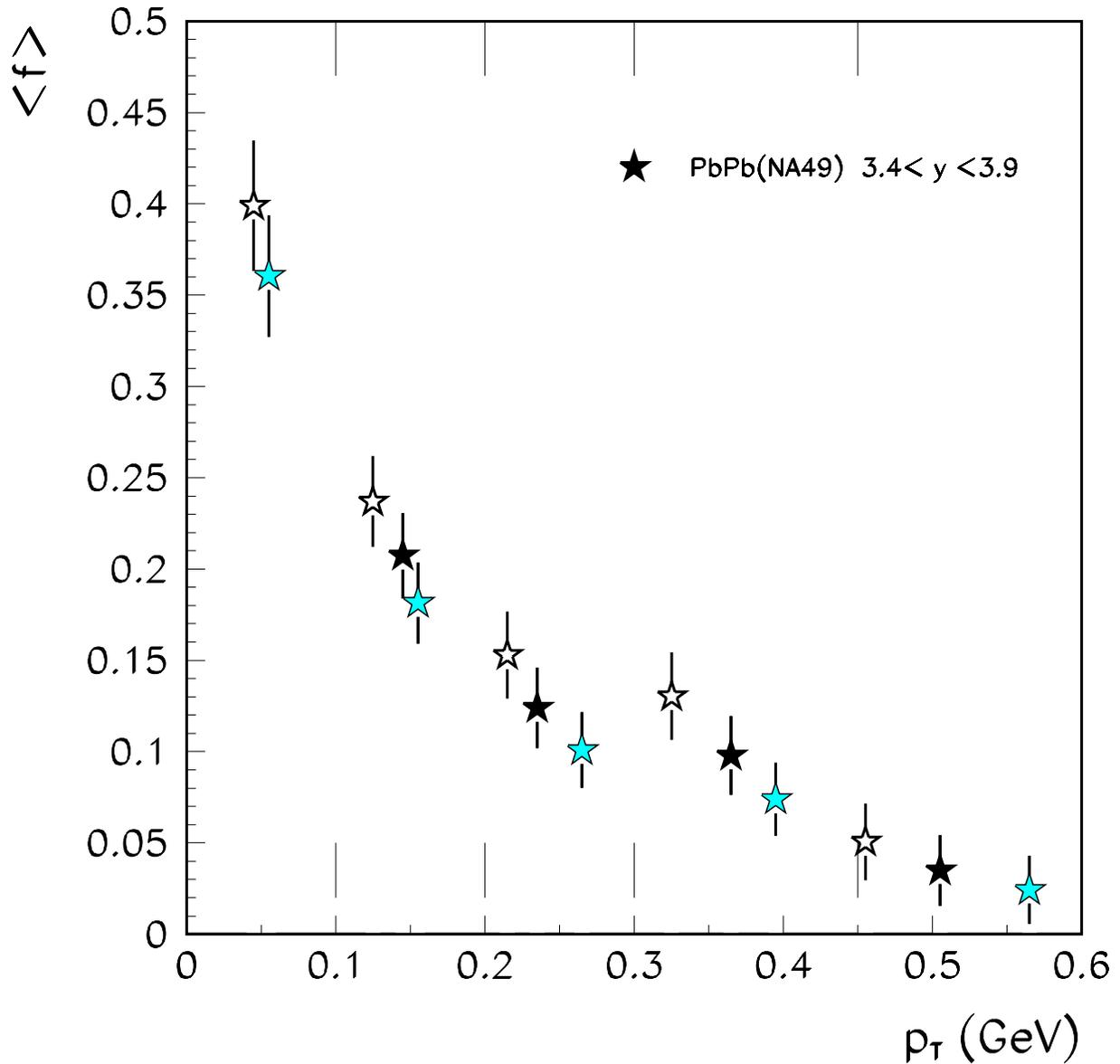,bbllx=0,bblly=0,bburx=500,bbury=500,width=16cm}
\caption{
Each data point was placed at three different places in $p_T$: at the
average value $\bar p_T$ of the corresponding bin (solid), at 
$0.9\,\bar p_T$ (open), and at $1.1\,\bar p_T$ (grey symbols). 
The bin positions are indicated by larger tics. The resulting 
smearing follows the general shape of the distribution. 
}
\label{FPP1}
\end{figure}

One should also specify the $p_T$ value representative for a given
$p_T$-interval, both in order to calculate $\langle f \rangle$ and
to allow for a meaningful model comparison, i.e. to properly place the
calculated points in the plot $\langle f \rangle (p_T)$. Fortunately, 
the result is almost insensitive to systematical errors in the choice
of the representative $p_T$ value. To estimate the effect of this
error we have shifted each measured point by 10\% up and down in $p_T$
from the average $p_T$ for the interval, as shown in Fig.~\ref{FPP1}. 
The resulting smearing is elongated along the actual shape of the
entire distribution. The systematical error in the choice of the
$p_T$ value representative for a given $p_T$ interval is 
therefore essentially harmless. 

With the S-nucleus collision data from NA35 and NA44 and with the
$\pi$-p collision data from NA22 we had yet another problem: the
two-pion correlation functions of these data have not been fitted with
the proper functional form which includes the out-longitudinal cross
term $R_{ol}$~\cite{cross-term}. For those data the resulting volume
term in the denominator of Eq.~(\ref{5}) reduces to $R_s R_o R_l$. 
To estimate the systematic error arising from the omission of the 
cross term we considered the ratio of the expressions for 
$\langle f \rangle$ with and without the cross term:
 \begin{equation}
 \label{15}
   \frac{\langle f \rangle} {\langle f \rangle (R_{ol}=0)} = 
   \frac{1}{\sqrt{1 - \frac{R^4_{ol}} {R^2_o R^2_l}}}.
 \end{equation}
With the explicit expressions given in \cite{cross-term} and the Cauchy 
inequality one shows that $R_{ol}^4 < R^2_o R^2_l$ such that this ratio 
is always well-defined. It must be 1 at midrapidity since there 
$R_{ol}=0$ by symmetry. Larger values of 1.2-1.3 were only found 
for data far away from mid-rapidity for which the measured cross term 
was comparable to $R_o$ or $R_l$. In this publication we have considered 
data for which no cross term was included in the analysis only near 
mid-rapidity where $R_{ol}$ is small. We estimate that that this 
procedure limits the (downward) systematic error on  
$\langle f \rangle$ due to this effect to less than 5\%.  

\vskip 0.3cm

\begin{figure}[*]
\epsfig{file=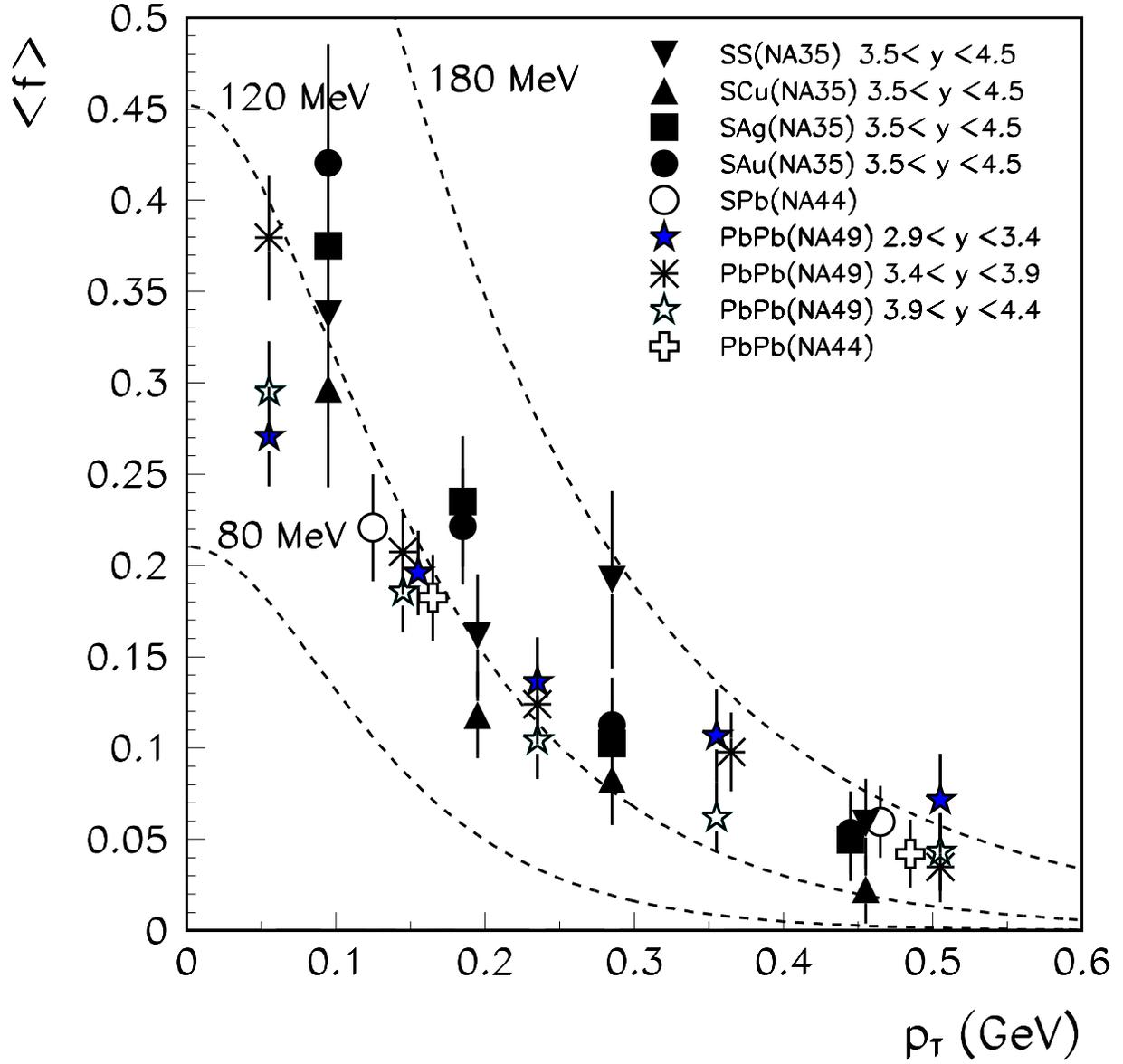,bbllx=0,bblly=0,bburx=500,bbury=500,width=16cm}
\caption{
Phase-space density as a function of $p_T$. Different heavy-ion data
sets from the SPS are indistinguishably similar. Dashed lines indicate 
Bose-Einstein distributions for three choices of the local freeze-out
temperature: 80 MeV, 120 MeV and 180 MeV.
}
\label{FPP2}
\end{figure}

\begin{figure}[*]
\epsfig{file=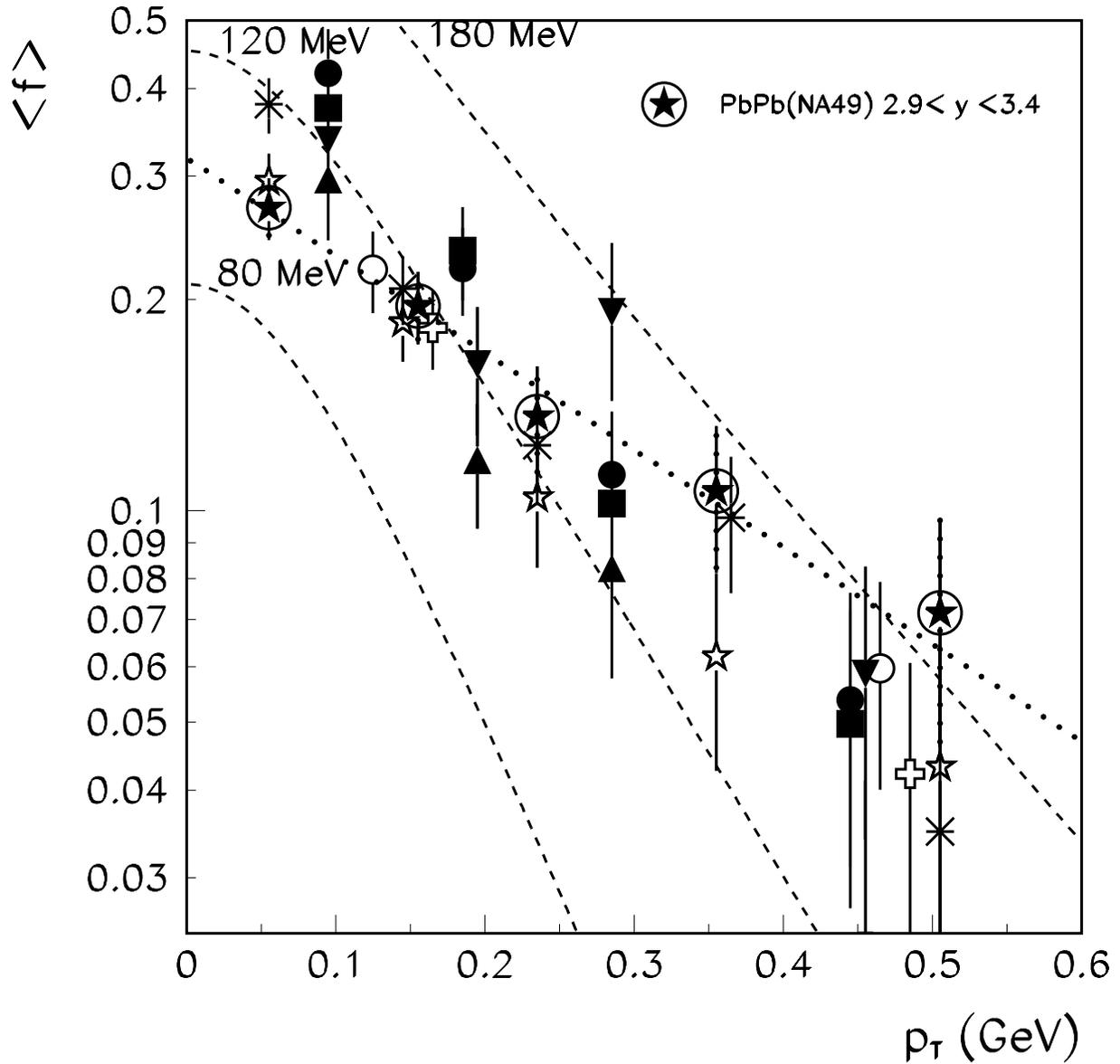,bbllx=0,bblly=0,bburx=500,bbury=500,width=16cm}
\caption{
Logarithmic representation of the data shown in Fig.\ref{FPP2}.
Discrepancies between the data and the Bose-Einstein distribution
($T=$120 MeV) can be explained by radial transverse flow. The dotted
line represents an exponential fit to the mid-rapidity Pb-Pb data (see 
text).
}
\label{FPP3}
\end{figure}

\begin{figure}[*]
\epsfig{file=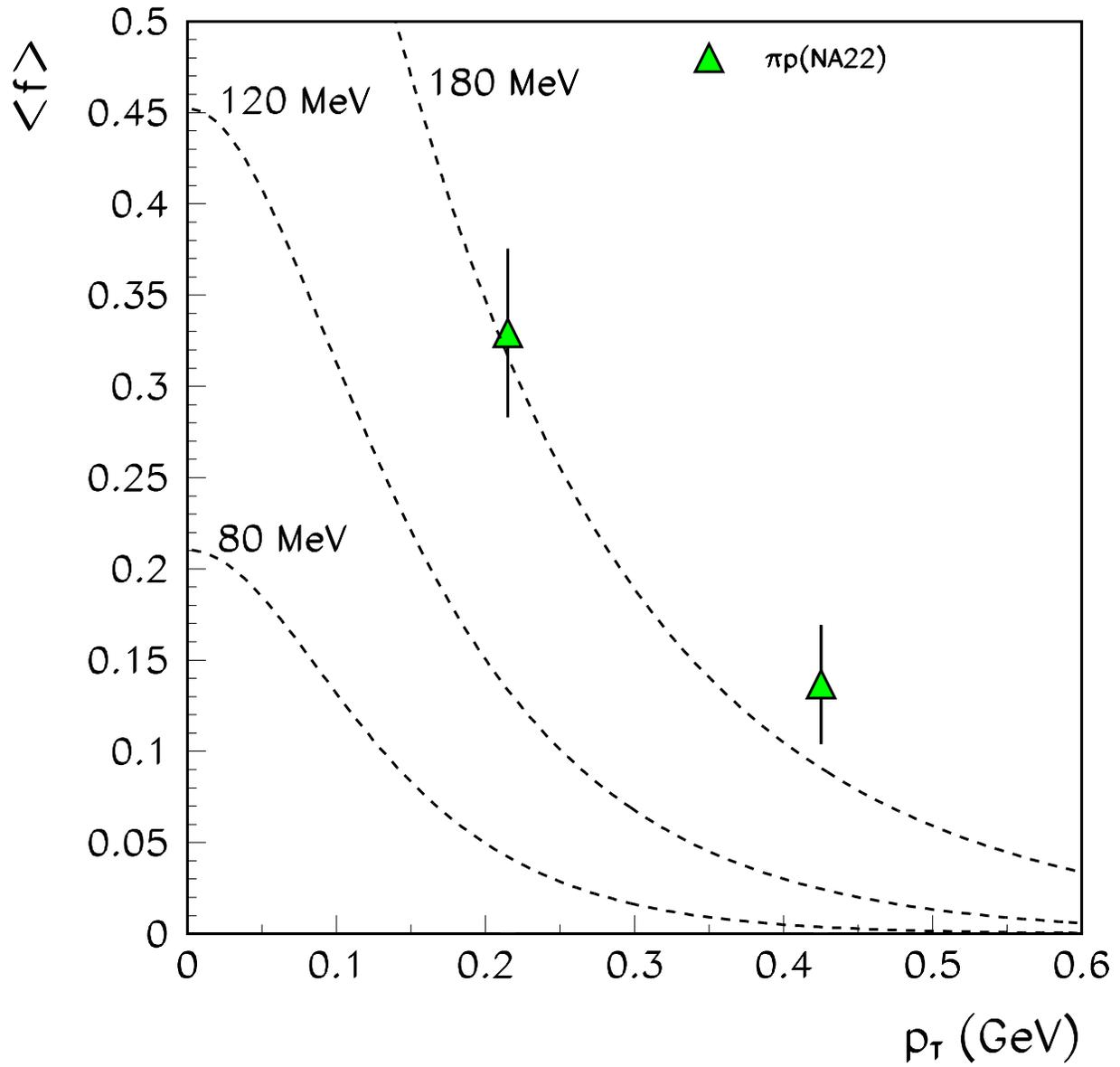,bbllx=0,bblly=0,bburx=500,bbury=500,width=16cm}
\caption{
NA22 $\pi$-p data. They are consistent with a Bose-Einstein distribution
of temperature $T=180$ MeV. 
}
\label{FPP2NA22}
\end{figure}

\begin{figure}[*]
\epsfig{file=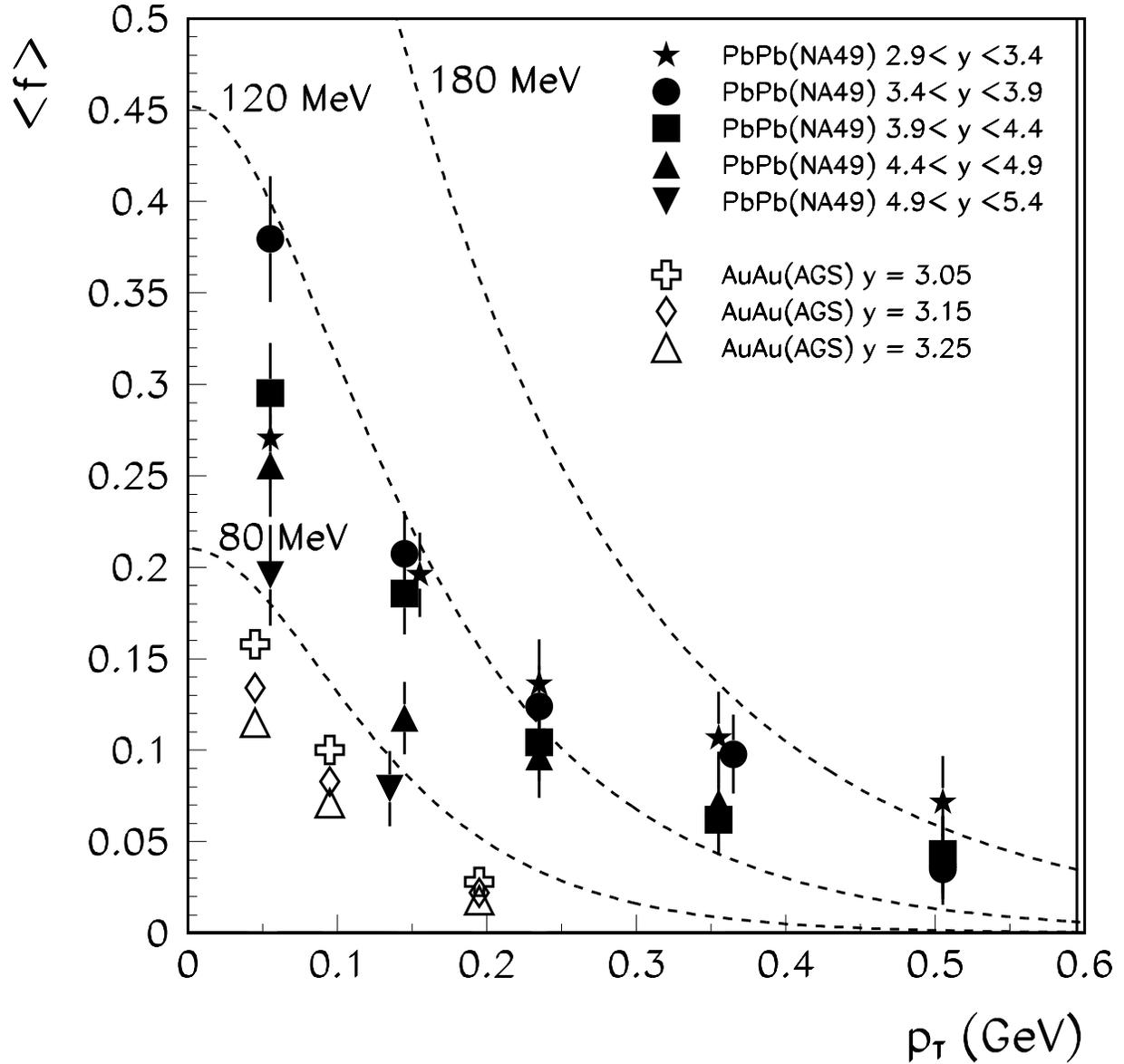,bbllx=0,bblly=0,bburx=500,bbury=500,width=16cm}
\caption{
 Average freeze-out phase-space density for A+A collisions ($A\approx
 200$) as a function of $p_T$ in different rapidity intervals.
 At very forward rapidities $\langle f \rangle$ decreases both at the
 SPS and the AGS as expected from a thermalized, longitudinally
 expanding system (see text). 
}
\label{FPP6}
\end{figure}

{\it Results.}
In Figures \ref{FPP2}-\ref{FPP6} we show the average phase-space
density as a function of $p_T$, for different systems and rapidity 
bins, as extracted from Eq.~(\ref{5}). The error bars reflect the
statistical errors of the single-particle spectra and correlation
radii, but not the systematic uncertainties discussed above (including 
the dominant uncertainty in the intercept $\lambda$). We also plot for 
comparison the Bose-Einstein distribution (\ref{1}) for a static
system ($u(x)=0$) and different temperatures $T$. In doing so we set
$E\approx \sqrt{m^2+p_T^2}$ which is justified if the longitudinal
pair momentum can be neglected. Since the correlation data which we
used were analyzed in the so-called ``fixed longitudinal co-moving
system'' (FLCMS) in which the pair rapidity corresponding to the
center of the rapidity bin vanishes, this approximation holds as long
as the considered rapidity interval is narrow. Rapidity intervals of
0.5 (as in the NA49 Pb-Pb data) and 1 (as for the NA35 S-nucleus data) 
result in a systematic overestimate of the theoretical distribution
function by approximately 2\% and 7\%, respectively.

From the results presented in Figures~\ref{FPP2}--\ref{FPP6}, 
one may draw the following conclusions:
{\vskip 0.15 cm}

\begin{enumerate}

\item
{\bf Universal average phase-space density at freeze-out:}\\
Even though the heavy-ion data span about an order of magnitude in 
multiplicity density ($dn^-/dy$(S-S)=22, $dn^-/dy$(S-Cu)=24,
$dn^-/dy$(S-Ag)=28, $dn^-/dy$(S-Au)=37, and $dn^-/dy$(Pb-Pb)=40-185,
depending on rapidity), the resulting average phase-space densities 
vary by much less. Given the large error bars, all the nuclear collision 
data at mid-rapidity from the SPS experiments in 
Figs.~\ref{FPP2},\ref{FPP3} are almost indistinguishable. Freeze-out 
happens in all cases at comparable values of the average phase-space 
density.

\item
{\bf Rough agreement with the Bose-Einstein distribution:}\\
The dashed lines in Fig.~\ref{FPP2} indicate thermal Bose-Einstein 
distributions for static sources at various freeze-out temperatures. A 
rough comparison with the SPS heavy-ion data indicates consistency 
with temperatures in the range of 100-140 MeV. Thermal freeze-out 
temperatures in this domain were recently obtained in analyses of 
the measured spectra and correlation functions \cite{NA49,CN96,QM97,T99}.

\item
{\bf Multiparticle symmetrization effects are small:}\\
Irrespective of the collision system and $p_T$, we find phase-space
densities smaller than 0.5. For $\langle f \rangle \ll 1$ the 
Bose-Einstein phase-space enhancement is dominated by two-particle 
symmetrization effects, and multiparticle symmetrization effects are 
weak \cite{WH99}. This is an important consistency check for the 
current practice of calculating the two-particle correlation function 
from the two-particle symmetrized contributions only. For 
$\langle f \rangle  < 0.5$ the system is far away from a pion laser.
We find no sign for a striking pion excess or a pion condensate.
 
\item
{\bf Radial flow:}\\
Looking in more detail, in particular applying a logarithmic scale as
in Fig.~\ref{FPP3}, one finds that the data indicate a somewhat slower
decrease with increasing $p_T$ than the Bose-Einstein curve. Particularly 
strong differences are seen for the NA49 Pb-Pb data at mid-rapidity 
($2.9<y<3.4$). These data can be fitted with the function 
$\exp[-1.1(1)-p_T/0.31(6)\,{\rm GeV}]$ ($\chi^2$/ndf=0.2), presented 
in Fig.~\ref{FPP3} by a dotted line, which has a considerably flatter 
slope than the Bose-Einstein distribution. This behaviour can be 
reproduced by a model for the emission function which includes radial 
collective expansion and whose parameters are adjusted to reproduce 
the single- and two-particle spectra \cite{T99}. The (strong) radial 
expansion adds extra transverse momentum to the particles (Doppler 
blue shift), i.e. the local $\langle f \rangle$ values appear in the 
lab frame at a higher $p_T$ than in the rest frame of the effective 
source. A detailed study will be published elsewhere \cite{T99}.

\item
{\bf Decoupling at high temperature in $\pi$-p collisions:}\\
In contrast to freeze-out in nuclear collisions which takes place in
two steps (chemical freeze-out of particle abundances at around
$T_{\rm chem} \simeq 170-180$~MeV~\cite{Becattini}, thermal freeze-out
of momentum spectra at around $T_{\rm therm} \simeq 120$
MeV~\cite{NA49}), pion production in $\pi$-p collisions~\cite{NA22} is
essentially immediate, without the second evolution stage, and
therefore common chemical and thermal freeze-out temperatures of
around 170-180 MeV should be expected. The data on $\langle f\rangle$
are indeed consistent with this expectation, as seen in
Fig.~\ref{FPP2NA22}. 

\item
{\bf Rapidity dependence:}\\
A certain departure from the universal scaling is seen for the data at
rapidities close to the projectile rapidity, see Fig.~\ref{FPP6}, both
at AGS and SPS; but again the results at these two widely different
beam energies are mutually consistent and agree with the expectations
from a thermalized expanding source. The sources at AGS and SPS
energies show strong longitudinal expansion at freeze-out, but with
maximal longitudinal flow rapidities well below the beam rapidity
($\eta_{\rm flow,max} \approx 1.7$ at the SPS and $\approx 1.1$ at the
AGS \cite{SH93,Stachel}). Pions with CMS rapidities larger than
$\eta_{\rm flow,max}$ thus come from the tail of the thermal {\em
  longitudinal} momentum distribution \cite{AGS}. A similar decrease 
of $\langle f\rangle$ which is seen near midrapidity as a function of 
$p_T$ is thus seen near beam rapidities as a function of $y$.

\end{enumerate}

\vskip 0.3cm

{\it Acknowledgement:} This work was initiated during the Heavy Ion 
Workshop at the Institute for Nuclear Theory (Seattle) in March 1998, 
and U.H. would like to thank the INT for its hospitality. We also 
acknowledge support by DAAD, DFG, GSI, BMBF, and the US Department of Energy.

\end{document}